\documentclass[]{spie}

\usepackage{amsmath}
\DeclareMathOperator*{\argmax}{arg\,max}
\DeclareMathOperator*{\argmin}{arg\,min}
\usepackage{amsmath,amsfonts,amssymb}
\usepackage{graphicx}
\usepackage{xcolor}
\usepackage[colorlinks=true, allcolors=blue]{hyperref}
\usepackage{multirow}
\usepackage{graphicx,amsmath,gensymb}
\usepackage[nobysame,lite,short-months]{amsrefs}

\title{On design of hybrid diffractive optics for achromatic extended depth-of-field (EDoF) RGB imaging }

\author[a]{Seyyed Reza Miri Rostami}
\author[a]{Samuel Pinilla}
\author[a]{Igor Shevkunov}
\author[a]{Vladimir Katkovnik}
\author[a]{Karen Egiazarian}

\affil[a]{Computing Sciences Unit, Faculty of Information Technology and Communication Sciences, Tampere University, FI-33720 Tampere, Finland}

\authorinfo{Further author information: (Send correspondence to A.A.A.)\\A.A.A.: E-mail: SeyyedReza.MiriRostami@tuni.fi, Telephone: +358 46 589 0487}

\begin{document}
\maketitle

\begin{abstract}
A hybrid imaging system is a simultaneous physical arrangement of a refractive lens and a multilevel phase mask (MPM) as a diffractive optical element (DOE). The favorable properties of the hybrid setup are improved extended-depth-of-field (EDoF) imaging and low chromatic aberrations. We built a fully differentiable image formation model in order to use neural network techniques to optimize imaging. At the first stage, the design framework relies on the model-based approach with numerical simulation and end-to-end joint optimization of both MPM and imaging algorithms. In the second stage, MPM is fixed as found at the first stage, and the image processing is optimized experimentally using the CNN learning-based approach with MPM implemented by a spatial light modulator. The paper is  concentrated on a comparative analysis of imaging accuracy and quality for design with various basic optical parameters: aperture size, lens focal length, and distance between MPM and sensor. We point out that the varying aperture size, lens focal length, and distance between MPM and sensor are for the first time considered for end-to-end optimization of EDoF. We numerically and experimentally compare the designs for visible wavelength interval [400-700]~nm and the following EDoF ranges: [0.5-100]~m for simulations and [0.5-1.9]~m for experimental tests. This study concerns an application of hybrid optics for compact cameras with aperture [5-9] mm and distance between MPM and sensor [3-10] mm.
\end{abstract}


\maketitle
\section{Introduction}
\label{sec:introduction}
End-to-end optimization of diffractive optical element (DOE) profile (e.g., binary/multi-level phase elements \cite{leveque2020co,baek2020end,10.1117/1.OE.60.5.051204}; meta-optical elements included \cite{chen2020flat,Tseng2021NeuralNanoOptics,BayatiPestourieColburnLinJohnsonMajumdar+2021,doi:10.1126/sciadv.aar2114,Whitehead:22}) has gained an increasing attention in emerging applications such as photography \cite{sitzmann2018end,dun2020learned}, augmented reality \cite{krajancich2020factored}, spectral imaging \cite{10.1145/3306346.3322946}, microscopy \cite{adams2017single}, among others that are leading the need for highly miniaturized optical systems \cite{antipa2018diffusercam,yanny2019miniature}, etc. The design methodology is performed by building numerical differentiable models for propagation of light fields through the physical setup in order to  employ for modeling and optimization neural networks methods. In particular, the power-balanced diffractive hybrid optics (lens and MPM) is proposed and studied in \cite{Rostami:21}, the methodology that is intended to be followed in this work, where a spatial light modulator (SLM) is used in experiments for implementation of MPM encoding of light fields. 

In this work, the elements of interest to be jointly designed are MPM and image processing algorithms. The techniques and algorithms used for this design take advantage of those developed in \cite{Rostami:21,https://doi.org/10.48550/arxiv.2203.16407}. As in \cite{Rostami:21}, the targeted imaging problem is Extended Depth-of-Field (EDoF) with reduced chromatic aberrations. We exploit a fully differentiable image formation model for joint optimization of optical and imaging parameters for the designed computational camera using neural networks. In particular, for the number of levels and Fresnel order features, we introduce a smoothing function because both parameters are modeled as piecewise continuous operations. The paper is concentrated on pragmatical aspects of the design, especially, on the imaging quality and accuracy as functions of basic optical parameters: aperture size, lens focal length, thickness of MPM, distance between MPM and sensor, $F$-number. We numerically and experimentally compare the designed systems for visible wavelength interval $(400-700)$~nm and depth-of-ﬁeld range defined as ($0.5$-$100$)~m for numerical and (0.5-2)~m for experimental tests. The study concerns application of the hybrid optics for compact cameras with aperture $(5 - 9)$~mm and lens focal length $(3 - 10)$~mm. We point out that the variables aperture size, lens focal length, and distance between MPM and sensor are for the first time considered for end-to-end optimization of EDoF.

The contribution of this work can be summarized as follows.
\begin{itemize}
	\item End-to-end optimization methodology for the joint design of MPM in the hybrid optics and imaging algorithms, showing high efficiency in terms of image accuracy and visual quality. \vspace{-0.5em}
	\item Optimal hybrid setup in terms of the optimal balance between aperture size and lens focal length concluded from multiple simulated experiments.	\vspace{-0.5em}
	\item Algorithms for using SLM as MPM in the hybrid optics with learning-based CNN optimization of inverse imaging.\vspace{-0.5em}
	\item The advanced achromatic EDoF imaging of the designed system as compared with conventional compound multi-lens cameras such as in iPhone Xs Max.
\end{itemize}

\section{End-to-End Optimization of Imaging with Hybrid optics} 
\label{Power-Balanced-math}

\begin{figure}[t!]
	\centering
	\includegraphics[width=0.6\linewidth]{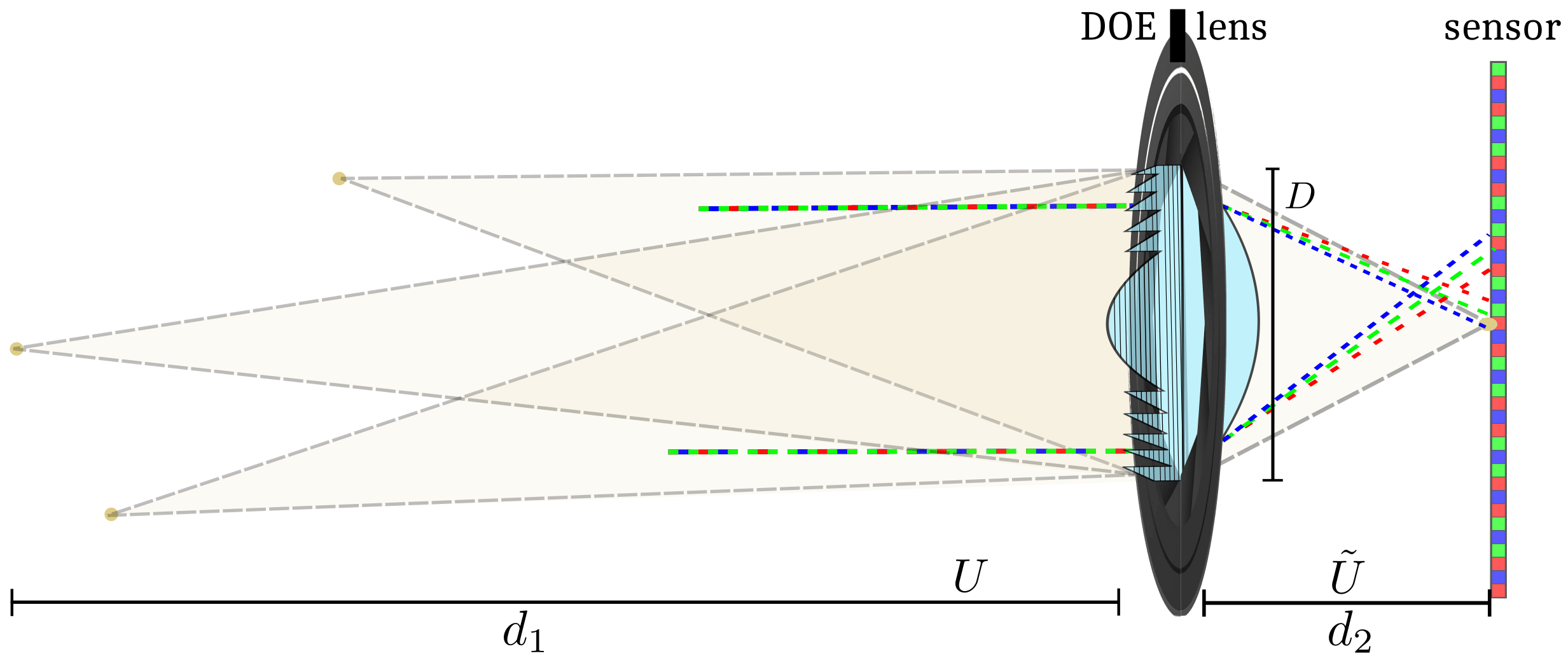}
	\caption{ A light wave with a given wavelength and a curvature for a point source at a distance $d_{1}$ propagates to the aperture plane containing MPM (refractive index $n$) to be designed. The MPM modulates the phase of the incident wavefront. The resulting wavefront propagates through the lens to the aperture-sensor, distance $d_{2}$, via the Fresnel propagation model. The intensities of the sensor-incident wavefront define PSFs of the diffractive hybrid optical system.}
	\label{fig:system}
\end{figure}

The optical setup of the imaging system is depicted in Figure~\ref{fig:system}, object, aperture, and sensor are $2D$ flat, where $d_{1}$ is a distance between the object and the aperture, $d_{2}$ is a distance from the aperture to the sensor ($d_{2}\ll d_{1}$), $f_{\lambda_{0}}$ is a lens focal length. In what follows, we use coordinates $(x,y)$, and $(u,v)$ for aperture, and sensor planes, respectively. In this section, we mainly follow the image formation modeling and design optimization presented in  \cite{Rostami:21}. These results are included for completeness of presentation and in order to give a clear picture of our approach, methodology, and algorithms.
\vspace{-0.6em}

\subsection{Image Formation Model}
\subsubsection{PSF-based RGB imaging}
Based on the Fresnel diffraction wavefront propagation, the response of an optical system to an input wavefront is modeled as a convolution of the system's PSF and a true object-image. Let us assume that there are both a lens and MPM in the aperture, then a generalized pupil function of the system for intensity imaging shown in Figure~\ref{fig:system} is of the form (see Eqs. (5-23)-(5-28) in \cite{goodman2005introduction})
	\begin{equation}
	\mathcal{P}_{\lambda}(x,y)=\mathcal{P}_{A}(x,y)e^ {\frac{j\pi }{\lambda }\left( {\frac{1}{d_{1}}}+{\frac{1}{d_{2}}-}\frac{1}{f_{\lambda }}\right) \left( x^{2}+y^{2}\right) +j\varphi_{\lambda _{0},\lambda}(x,y)}.
	\label{GPF}
	\end{equation}		
	
	
In \eqref{GPF}, $f_{\lambda}$ is a lens focal length for the wavelength $\lambda $, $P_{A}(x,y)$ represents the aperture of the optics and $\varphi_{\lambda _{0},\lambda }(x,y)$ models the phase delay enabled by MPM for the wavelength $\lambda $ provided that $\lambda _{0}$ is the wavelength design-parameter for MPM. In this formula, the phase $\frac{j\pi }{\lambda }\left( \frac{1}{d_{1}}+\frac{1}{d_{2}}\right) \left(x^{2}+y^{2}\right) $ appears due to propagation of the coherent wavefront from the object to the aperture (distance $d_{1}$) and from the aperture to the sensor plane (distance $d_{2}$), and $\frac{-j\pi }{\lambda f_{\lambda }} \left(x^{2}+y^{2}\right) $ is a quadratic phase delay due to the lens. 
For the lensless system 
\begin{equation}
	\mathcal{P}_{\lambda}(x,y)=\mathcal{P}_{A}(x,y)e^{\frac{j\pi }{\lambda }\left( {\frac{1}{d_{1}}}+{\frac{1}{d_{2}}}\right) \left(x^{2}+y^{2}\right) +j \varphi_{\lambda _{0},\lambda }(x,y)},
	\label{GPF-lensless}
	\end{equation}
and for the lens system without MPM, $\varphi_{\lambda _{0},\lambda}(x,y)\equiv 0$ in \eqref{GPF}.

In the hybrid system, which is the topic of this paper, the generalized aperture takes the form
\begin{equation}
	\mathcal{P}_{\lambda}(x,y)=\mathcal{P}_{A}(x,y)e^{\frac{j\pi }{\lambda }\left({\frac{1}{d_{1}}}+{\frac{1}{d_{2}}}-\frac{1}{f_{\lambda }} \right) \left( x^{2}+y^{2}\right) +j\varphi_{\lambda _{0},\lambda, \alpha }(x,y)},
	\label{GPF-hybrid}
\end{equation}
where the optical power of the hybrid is shared between the lens with the optical power $1/f_{\lambda}$ and the MPM due to the quadratic phase component included in the phase delay of MPM. The magnitude of the latter phase is controlled by a real-valued parameter $\alpha$. 

The PSF of the coherent monochromatic optical system for the wavelength $\lambda$ is calculated by the formula~\cite{goodman2005introduction}
\begin{equation}
	PSF_{\lambda }^{coh}(u,v)=\mathcal{F}_{\mathcal{P}_{\lambda}}\left(\frac{u}{d_{2}\lambda },\frac{v}{d_{2}\lambda }\right),
	\label{PSF=COHERENT} 
	\end{equation}
where $\mathcal{F}_{\mathcal{P}_{\lambda}}$ is the Fourier transform of $\mathcal{P}_{\lambda }(x,y)$. Then, PSF for the corresponding incoherent imaging, which is a topic of this paper, is a squared absolute value of $PSF_{\lambda }^{coh}(u,v)$. After normalization, this PSF function takes the form
	\begin{equation}
	PSF_{\lambda }(u,v)= \frac{\left \lvert PSF_{\lambda }^{coh}(u,v)\right\rvert^{2}}{\iint_{-\infty }^{\infty}\left \lvert PSF_{\lambda }^{coh}(u,v)\right\rvert ^{2}dudv}. 
	\label{PSF}
\end{equation}
	
We calculate PSF for RGB color imaging assuming that the incoherent radiation is broadband and the intensity registered by an RGB sensor per $c$-band channel is an integration of the monochromatic intensity over the wavelength range $\Lambda $ with the weights $T_{c}(\lambda )$ defined by the sensor color filter array (CFA) and spectral response of the sensor. Normalizing these sensitivities on $\lambda$, i.e. $\int_{\Lambda }T_{c}(\lambda )d\lambda=1$, we obtain RGB channels PSFs
\begin{align}
	PSF_{c}(u,v) = \frac{\int_{\Lambda }PSF_{\lambda }(u,v)T_{c}(\lambda )d\lambda }{\iint_{-\infty }^{\infty}\int_{\Lambda }PSF_{\lambda }(u,v)T_{c}(\lambda )d\lambda dudv}
	\text{, } c\in \{r,g,b\}\text{,}
	\label{PSF_ave} 
\end{align}
where the monochromatic $PSF_{\lambda}$ is averaged over $\lambda$ with the weights $T_{c}(\lambda)$. 
	
Thus, for PSF-based RGB imaging, we take into consideration the spectral properties of the sensor and in this way obtain accurate modeling of image formation \cite{katkovnik2019lensless}. The OTF for \eqref{PSF_ave} is calculated as the Fourier transform of $PSF_{c }(u,v):$
\begin{align}	
	OTF_{c}(f_{x},f_{y})=\iint_{-\infty }^{\infty }PSF_{c}(u,v)e^{-j2\pi (f_{x}u+f_{y}v)}dudv,
 \label{OTF1}
\end{align}
where $(f_{x},f_{y})$ are the Fourier frequency variables.

\subsubsection{From PSFs to Imaging}
Let us introduce $PSFs$ for defocus scenarios with notation $PSF_{c,\delta }(x,y)$, where $\delta$ is a defocus distance in $d_{1}$, such that $d_{1}=d_{1}^0+\delta$ with $d_{1}^0$ equal to the focal distance between the aperture and the object. Introduce a set $\mathcal{D}$ of defocus values $\delta \in\mathcal{D}$ defining area of the desirable EDoF. It is worth noting that the corresponding optical transfer functions are used with notation $OTF_{c,\delta}(f_{x},f_{y})$. The definition of $OTF_{c,\delta}(f_{x},f_{y})$ corresponds to \eqref{OTF1}, where $PSF_{c}$ is replaced by $PSF_{c,\delta}$. Thus, let $I_{c, \delta }^s(u,v)$ and $I_{c}^o(u,v)$ be wavefront intensities at the sensor (registered focused/misfocused images) and the intensity of the object (true image), respectively. Then, $I_{c, \delta}^s(u,v)$ are obtained by convolving the true object-image $I_{c}^o(u,v)$ with $PSF_{c,\delta}(u,v)$ forming the set of misfocused (blurred) color images
\begin{equation}
I_{c,\delta}^{s}(x,y)=PSF_{c,\delta}(x,y)\circledast I_{c}^{o}(x,y),
\label{1}
\end{equation}
where $\circledast$ stays for convolution. In the Fourier domain we have
\begin{equation}
I_{c,\delta}^{s}(f_{x},f_{y})=OTF_{c,\delta}(f_{x},f_{y}) \cdot I_{c}^{o}%
(f_{x},f_{y}).
\label{2}
\end{equation}
The indexes $(o,s)$ stay for object and sensor, respectively.

\subsubsection{EDoF Image Reconstruction} \label{recon}
For image reconstruction from the blurred data $\{I_{c,\delta}^{s,k}(f_{x},f_{y})\}$, we use a linear filter with the transfer function $H_{c}$ which is the same for any defocus $\delta\in\mathcal{D}$. We formulate the design of the inverse imaging transfer function $H_{c}$ as an optimization problem 

\begin{equation}
\hat{H}_{c} \in \argmin_{H_{c}} \hspace{.5em}\underbrace{\frac{1}{\sigma^{2}}\sum_{\delta,k,c} \omega_{\delta}||I_{c}^{o,k}-H_{c}\cdot I_{c,\delta}^{s,k}||_{2}^{2}+\frac{1}{\gamma}\sum_{c}||H_{c}||_{2}^{2}}_{J}, 
\label{22_}
\end{equation}
where $k\in K$ stays for different images, $I_{c}^{o,k}$ and $I_{c,\delta}^{s,k}$ are sets of the true and observed blurred images (Fourier transformed), $c$ for color, $\sigma^{2}$ stands for the variance of the noise, and $\gamma$ is a Tikhonov regularization parameter. The parameters $ \omega_{\delta}>0$ are the residual weights in \eqref{22_}. We calculate these weights as the exponential function $\omega_{\delta} = exp(-\mu\cdot |\delta|)$ with the parameter $\mu>0$.  The norm $||\cdot||_{2}^{2}$ is Euclidean defined in the Fourier domain for complex-valued variables.

Thus, we aimed to find $H_{c}$ such that the estimates $H_{c}\cdot I_{c,\delta}^{s,k}$ would be close to FT of the corresponding true images $I_{c}^{o,k}$. The second summand stays as a regularizer for $H_{c}$. Due to \eqref{2}, minimization on $H_{c}$ is straightforward leading to 
\begin{equation}
\hat{H}_{c}(f_{x},f_{y})=\frac{\displaystyle \sum_{\delta \in\mathcal{D}}\omega_{\delta}OTF_{c,\delta}^{\ast}(f_{x},f_{y})}{\displaystyle \sum_{\delta \in\mathcal{D}}\omega_{\delta}|OTF_{c,\delta}(f_{x},f_{y})|^{2}+\frac{reg}{\sum_{k}|I_{c}^{o,k}(f_{x},f_{y})|^{2}}},
\label{solH}
\end{equation}
where the regularization parameter $reg$ stays for the ratio $\sigma^{2}/{\gamma}$. 

Therefore, the reconstructed images are calculated as 
\begin{equation}
	\hat{I}_{c}^{o,k}(x,y) = \mathcal{F}^{-1}\{ \hat{H_{c}} \cdot I_{c,\delta}^{s,k} \},
\label{misfocus_color_data101}
\end{equation}
where $\mathcal{F}^{-1}$ models the inverse Fourier transform. For the exponential weight $\omega_{\delta} = exp(-\mu\cdot |\delta|)\text{, }\mu>0$ is a parameter that is optimized. The derived OTFs \eqref{solH} are optimal to make the estimates \eqref{misfocus_color_data101} efficient for all $\delta \in\mathcal{D}$, in this way, we are targeted on EDoF imaging.

\subsection{MPM Modeling and Design Parameters}
\label{MPM }
In our design of MPM, we follow the methodology proposed in \cite{katkovnik2019lensless}. The following parameters characterize the free-shape piece-wise invariant MPM: $h$ is a thickness of the varying part of the mask, $N$ is a number of levels, which may be of different height.

\subsubsection{Absolute Phase Model} The proposed absolute phase $\varphi_{\lambda _{0}, \alpha }$ for our MPM takes the form
\begin{align}
\varphi_{\lambda _{0}, \alpha }(x,y) = \frac{-\pi\alpha}{\lambda_0 f_{\lambda_{0}}} (x^{2} + y^{2}) + \beta(x^{3} + y^{3}) +\sum_{r=1,r\not =4}^{R}\rho_{r}P_{r}(x,y).
\label{abs=phase}
\end{align}
The factor with $\lambda_{0}$ in this equation is introduced for a proper scaling of the MPM's quadratic phase with the phase delay of the refractive lens. The parameter $\alpha$ in this factor  controls the optical power sharing between the lens and MPM. The cubic phase of a magnitude $\beta$ is a typical component for EDoF, the third group of the items is for parametric approximation of the free-shape MPM using the Zernike polynomials $P_{r}(x,y)$ with coefficients $\rho_{r}$ to be estimated. We exclude from this approximation the fourth Zernike  polynomial defining the quadratic defocus term because it is considered as the first item in $\varphi_{\lambda _{0}, \alpha }(x,y)$.

\subsubsection{Fresnel Order (thickness of MPM)} In radians, the mask thickness is defined as $Q=2\pi m_{Q}$, where $m_{Q}$ is called 'Fresnel order' of the mask which in general is not necessarily integer. The phase mask profile of the thickness $Q$ is calculated as
\begin{equation}
\hat{\varphi}_{\lambda _{0}, \alpha}(x,y) = mod(\varphi_{\lambda _{0}, \alpha }(x,y) + Q/2,Q)-Q/2. 
\label{lens4}
\end{equation}
The operation in \eqref{lens4} returns $\hat{\varphi}_{\lambda _{0}, \alpha}(x,y)$ taking the values in the interval $[-Q/2$, $Q/2)$. The parameter $m_{Q}$ is known as 'Fresnel order' of the mask. For $m_{Q}=1$ this restriction to the interval $[-\pi $, $\pi )$ corresponds to the standard phase wrapping operation.

\subsubsection{Number of Levels} 
The mask is defined on $2D$ grid $(X,Y)$ with the computational sampling period (computational pixel) $\Delta _{comp}$. We obtain a piece-wise invariant surface for MPM after the non-linear transformation of the absolute phase.
The uniform grid discretization of the wrapped phase profile $\hat{\varphi}_{\lambda _{0}, \alpha}(x,y)$ to the $N$ levels is performed as
\begin{equation}
	\theta_{\lambda _{0}, \alpha}(x,y) =\lfloor \hat{\varphi}_{\lambda _{0}, \alpha}(x,y) /N \rfloor \cdot N\text{,}
\label{lens5}
\end{equation}
where $\lfloor w \rfloor$ stays for the integer part of $w$. The values of $\theta_{\lambda _{0}, \alpha}(x,y)$ are restricted to the interval $[-Q/2$, $Q/2)$. $Q$ is an upper bound for thickness phase of $\theta_{\lambda _{0}, \alpha}(x,y)$. 

The introduced discretization and modulo functions are not differentiable, therefore we use a smoothing approximation to be able of optimizing the thickness and the number of levels of MPM by gradient descent algorithms. The details of this approximated function can be found in \cite{Rostami:21} .

The mask is designed for the wavelength $\lambda_{0}$. Thus, the piece-wise phase profile of MPM for the wavelength $\lambda$ is calculated as
\begin{equation}
\varphi_{MPM_{\lambda _{0},\lambda, \alpha}}(x,y) =\frac{\lambda _{0}(n(\lambda )-1)}{\lambda (n(\lambda _{o})-1)}\theta_{\lambda _{0}, \alpha}(x,y),
\label{BPM-final}
\end{equation}
where $\theta_{\lambda _{0}, \alpha}$ is the phase shift of the designed MPM and $n(\lambda)$ is the refractive index of the MPM material, $x\in X, y\in Y$. The MPM thickness $h$ in length units is of the form 
\begin{equation}
	h_{\lambda _{0}}(x,y)=\frac{\lambda _{0}}{(n(\lambda _{o})-1)} \frac{\theta_{\lambda _{0}, \alpha}} {2\pi }\text{.} 
	\label{lens8}
\end{equation}
	
\subsection{Optimization Framework}
\begin{figure}[t!]
	\centering
	\includegraphics[width=1\linewidth]{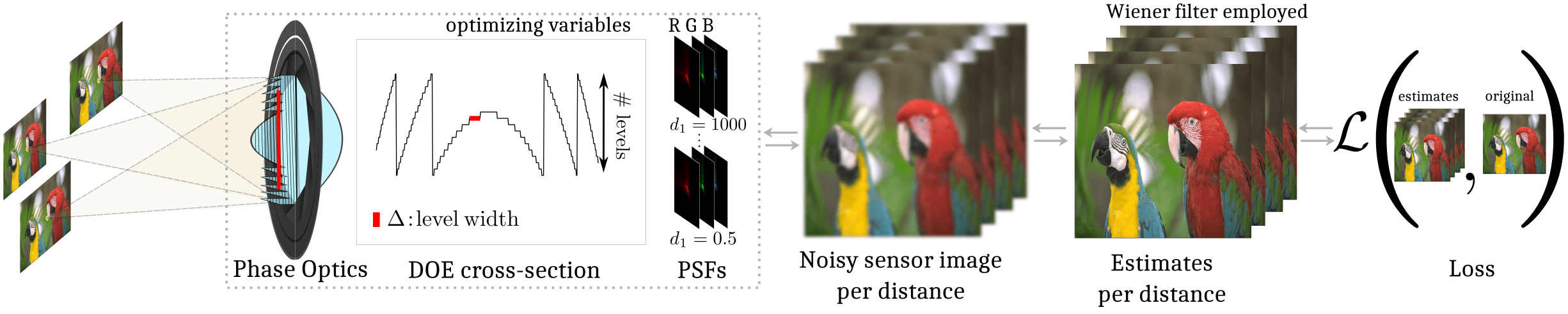}\vspace{-0.0em}
	\caption{ The optimal design framework of phase-encoded optics and image reconstruction algorithms for achromatic EDoF. The spectral PSFs are convolved with batches of RGB ground-truth images. The inverse imaging provides estimates of these images. Finally, a quality/accuracy loss $\mathcal{L}$, such as mean squared error with respect to the ground-truth images (or PSNR criterion), is defined on reconstructed images.}
	\label{fig:balancedSystem}
\end{figure}

The framework which is presented in Figure \ref{fig:balancedSystem} is developed to optimize the proposed optical system using the iterative NN algorithms with stochastic gradient ADAM optimizer. It can be downloaded from PyTorch with an optimized tensor library for Neural Network (NN) learning using GPUs \footnote{The Pytorch library \url{https://pytorch.org/}}. Some details concerning this framework are given in what follows in this section.

\subsubsection{Loss Function} Let $\Theta$ be a full set of the optimization parameters defined as
\begin{align}
\Theta= (\alpha,\beta,\rho_{r}, reg ).
\label{theta}
\end{align}
Then, we use the following multi-objective formulation of our optimization goals
\begin{equation}			
\hat{\Theta}=\argmax_{\Theta}(PSNR(\Theta,\delta), \delta\in\mathcal{D}).
\label{multiobjectivel}
\end{equation}%
In this formulation, we maximize all $PSNR(\Theta,\delta)$, $\delta\in\mathcal{D}$, simultaneously, i.e. to achieve the best accuracy for all focus and defocus situations. Here, $PSNR(\Theta,\delta)$ is calculated as the mean value of $PSNR^k(\Theta,\delta)$ over the set of the test-images, $k\in K$: 
\begin{equation}
	PSNR(\Theta,\delta)=mean_{k\in K}(PSNR^k(\Theta,\delta)).
	\label{PSNR_total}
\end{equation}

There are various formalized scalarization techniques reducing the multi-objective (vector) criterion to a scalar one. Usually, it is achieved by aggregation of multiple criteria in a single one (e.g. \cite{emmerich2018tutorial}). In this paper, we follow pragmatical heuristics comparing $PSNR(\hat\Theta,\delta)$ as the $1D$ functions of $\delta$ in order to maximize $PSNR(\Theta,\delta)$ for each $\delta\in\mathcal{D}$. Here, $\hat\Theta$ are estimates of the optimization parameter. In this heuristic, we follow the aim of the multi-objective optimization \eqref{multiobjectivel}. The key challenges in developing of the proposed optimization framework were to satisfy manufacturing constraints, finding stable optimization algorithms, and fitting models within memory limits.

\subsubsection{Parameters for MPM design and simulation tests} 
The sensor's parameters used in simulation correspond to the physical sensor used in our experiments: pixel size  $3.45~\mu m$ and resolution  $512\times 512$ pixels.
The Fourier transform for PSFs calculations are produced on the grid $3000\times 3000$ of the computational pixel size $\Delta_{comp}$=$2~\mu m$, defining discretization of lens and MPM. 
We fixed the number of MPM levels to $N=31$ and Fresnel order to $m_Q = 1$, the latter restricts the MPM phase wrapping to the interval [$-\pi$, $\pi$). The optimization stage includes finding the optimal $\alpha, \beta, \rho_{r}$ for the MPM design and $reg$ for the image inverse reconstruction using the Adam stochastic gradient descent solver with the step-size  $5\times 10^{-3}$.

We analyze and compare the hybrid optics of different lens diameter  (aperture size of hybrid) taking values $(5, 6, 7, 9) ~mm$ and lens focal length taking values $f=(3,5,7,10) ~mm$. The focus imaging distance for the hybrid is fixed to $d^0_{1}=1~m$. For each lens focal length $f$, $d_2$ is calculated according to the focusing equation $\frac{1}{d^0_{1}} + \frac{1}{d_{2}} = \frac{1}{f}$. These values of $d_2$ are very close to $f$.  It was concluded from our tests that $R=14$ (Zernike coefficients excluding the fourth polynomial in (\ref{abs=phase})) is enough and larger values of $R$ do not improve image quality significantly. The design wavelength is $\lambda_{0} = 510~nm$. An additive white Gaussian noise is included in observations with variance equal to $1\times 10^{-4}$. We choose 31  wavelengths, with step $10~nm$, covering the visual interval $(400-700)~nm$ to model RGB imaging. To enable EDoF imaging, we use Wiener filtering  with $d_{1} = 0.5, 0.6, 0.7, 1.0, 1.9, 10$, and $100.0 $ m. These $d_{1}$ define the defocus parameter $\delta$ in \eqref{solH} as $\delta=d_{1}-d^0_{1}$.  The optimization stage employs 200 epochs, which takes approximately 6 hours on NVIDIA GeForce RTX 3090 GPU with memory of 24GB.

\subsubsection{Data sets for optimization and tests} 
For optimization and training, we chose 1244 high-resolution RGB images from databases
\footnote{\url{https://data.vision.ee.ethz.ch/cvl/DIV2K/}, and \url{http://cv.snu.ac.kr/research/EDSR/Flickr2K.tar}.}.
For testing of the designed systems, we used 200 high-resolution RGB images from the same databases which are not included in the training set. In what follows, all illustrative materials (tables, curves, and images) are given for these test images.

\section{Simulation Tests}
\label{simulation-result}
In this section, we design the phase profiles for MPM in the hybrid optical setup with different aperture sizes (5, 6, 7, and 9) mm and lens focal lengths (3. 5, 7, and 10) mm. Our intention is to find combinations of these physical parameters for the best achromatic EDoF imaging. The corresponding numerical results obtained by simulation using the end-to-end joint optimization of optics and inverse imaging algorithms are presented in Table \ref{tab:optics_compare}.  The reported $PSNRs$ are averaged over $7$ depth (defocus) distances $d_{1}$  from the interval (0.5 - 100.0) m and over 200 RGB test-images. 

\begin{center} 
\begin{table}[ht]
\centering
\caption{ Comparative performance  of the hybrid optics: different lens diameter (aperture size) and lens focal length.} 

\label{tab:optics_compare}
\centering
\begin{tabular}{|c|c|c|c|c|c|c|c|}
\hline
 \multirow{2}{*}{Diameter (mm)} & \multirow{2}{*}{Focal length (mm)} & \multirow{2}{*}{$PSNR_{total}$(dB)} & \multicolumn{3}{c|}{PSNR per channel} & \multirow{2}{*}{\textit{F}-number} & \multirow{2}{*}{\shortstack{FOV \\ (degree)}} \\
 \cline{4-6}
 & & & R & G & B & & \\
\hline
 \multirow{4}{*}{5} & 3 & 31.48 & 28.43 &  34.61 & 31.64 & 0.6 & 99.3 \\
 & \textbf{5} & \textbf{41.58}& 43.20 & 44.65 & 39.82 & 1 & 70.5 \\
 & 7 & 38.75& 39.44 & 42.71 & 35.98 & 1.4 & 53.6 \\
 & 10 & 36.29& 36.98 & 40.11 & 31.84 & 2 & 38.9 \\
\hline
 \multirow{4}{*}{\textbf{6}} & 3 & 25.64 & 23.12 & 27.21 & 22.89 & 0.5 & 99.3 \\
 & \textbf{5} & \textbf{44.23}& \textbf{44.92} & \textbf{46.81} & \textbf{41.74} & \textbf{0.83} & \textbf{70.5} \\
 & 7 & 36.61& 39.41 & 40.87 & 30.29 & 1.17 & 53.6 \\
 & 10 & 33.66& 32.22 & 34.07 & 29.46 & 1.66 & 38.9 \\
\hline
 \multirow{4}{*}{7} & 3 & 25.8& 25.29 & 29.77 & 21.58 & 0.43 & 99.3 \\
 & 5 & 33.28& 29.09 & 37.09 & 29.47 & 0.71 & 70.5 \\
 & \textbf{7} & \textbf{36.14}& 34.61 & 39.26 & 29.93 & 1 & 53.6 \\
 & 10 & 31.65& 28.41 & 33.20 & 29.86 & 1.43 & 38.9 \\
\hline
 \multirow{4}{*}{9} & 3 & 24.21& 24.74 & 27.46 & 19.59 & 0.33 & 99.3 \\
 & 5 & 26.43& 22.45 & 28.40 & 26.05 & 0.54 & 70.5 \\
 & \textbf{7} & \textbf{34.79}& 29.08 & 39.10 & 30.19 & 0.76 & 53.6 \\
 & 10 & 30.97& 31.00 & 36.25 & 26.09 & 1.08 & 38.9 \\
\hline 
\end{tabular}
\end{table}
\end{center}

The imaging accuracy is evaluated and reported in two versions: $PSNR_{RGB}$ calculated for each of the color channels separately (column 4), and $PSNR_{total}$ calculated for all three color channels jointly (column 3). The best result (highest values of PSNR) is achieved by the setup with 6 mm aperture size and 5 mm lens focal length. These physical parameters result in  $F$-number=0.83 and 70.5-degree field of view (FOV). The $PSNR_{total}$ value for this case is equal to 44.23 dB, but it degrades dramatically for larger and smaller focal lengths within the fixed diameter. If we compare the PSNR for the color channels separately, the values for 6mm diameter designed hybrid optics are highest (all above 41 dB) and more or less the same for all color channels.

Note also, that for each lens diameter there is an optimal lens focal length and this optimal value is close to the diameter size. The optimal focal lengths for the diameters (5, 6, 7, and 9) mm are (5, 5, 7, and 7) mm, respectively. We may conclude that the lens focal length plays a crucial role in hybrid optics and there is a trade-off between imaging quality and FOV. Smaller focal length (in Table \ref{tab:optics_compare}, 3 mm) gives wider FOV at expense of less imaging accuracy. This conclusion is valid for all lens diameters in Table \ref{tab:optics_compare}.  

\begin{figure}[t!]
	\centering
	\includegraphics[width=0.6\linewidth]{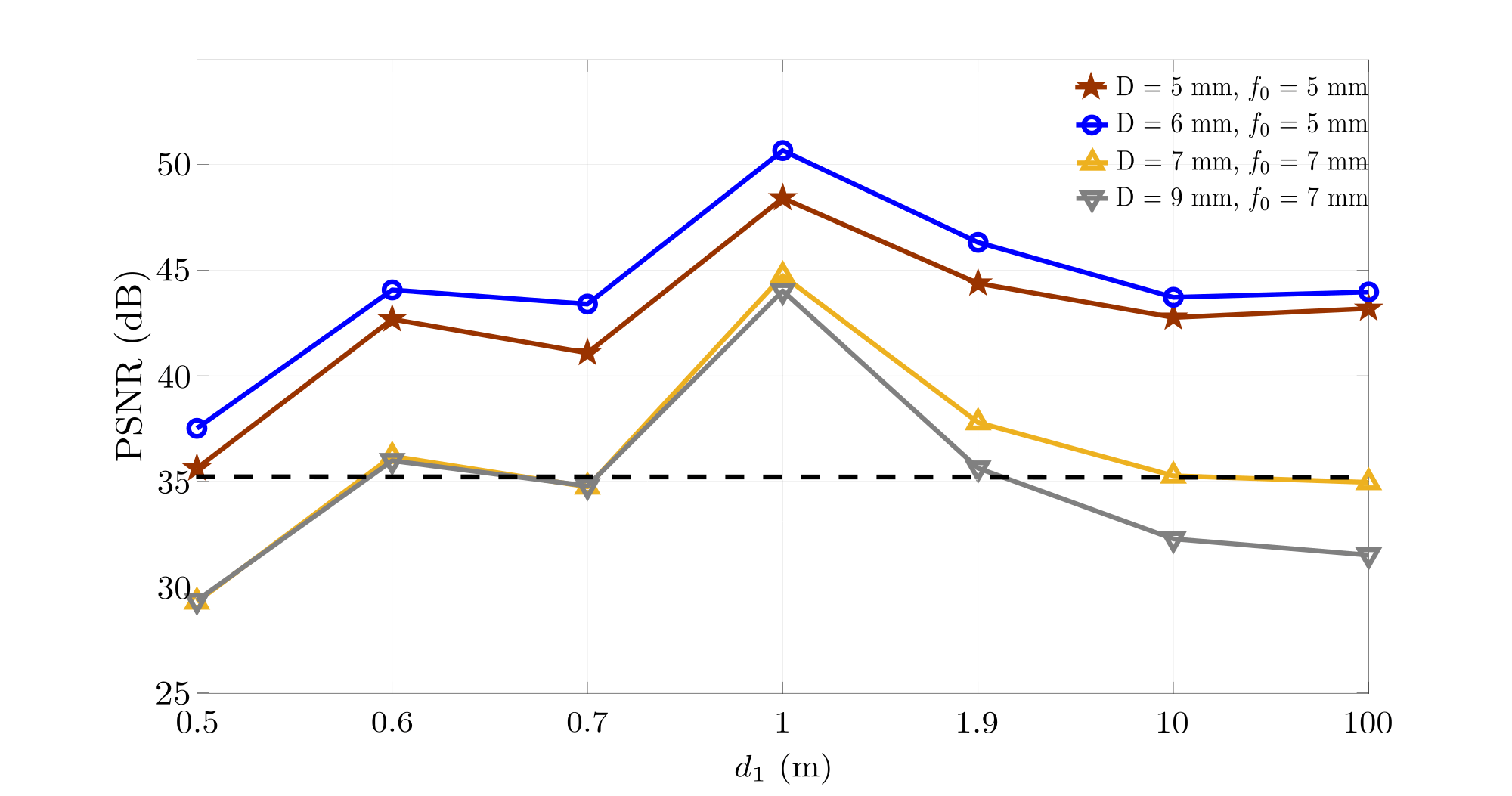}
	\caption{ PSNR curves of the optimized hybrid setups with 4 different aperture size D= (5, 6, 7, and 9) mm as a function of distance from the scene to the optics ($d_1$). The optimized hybrid setups with 5 and 6 mm diameters perform in the best way with more or less uniform PSNR values which are well above the good imaging quality line, PSNR = 35 dB, for all depths. The advantage of hybrid optics with $D=6$ mm versus $D=5$ mm is obvious of about 1 to 2 dB of PSNR values for each distance. The imaging with $D= (7 $ and $9)$ mm shows good results in the vicinity of the system focal point ($d_1 = 1 m$), but the performance is dropped for far and even quite close distances.}
	\label{fig:psnr_curve}
\end{figure}

Further information on the comparative performance of the imaging system with the optimized hybrid optics can be seen in Figure \ref{fig:psnr_curve}. Here we present PSNR curves as functions of $d_1$ (distance between the object and optics) averaged over 200 test images. The four curves are given for the four values of lens diameter with the corresponding optimal lens focal length as shown in Table \ref{tab:optics_compare}. 

The uniformly best performance is achieved by the 6 mm aperture hybrid optics with $f_{0} = 5 mm$. For this case, the PSNR value is about 37dB for the defocus point $d_{1}=0.5 m$. The peak of this curve is at $d_{1}=1.0 m$  with PSNR=50dB. Remind, that this is a focus point of the system. For larger defocus distances, $d_{1}>1$,  PSNR takes lower values which are nevertheless are close to 45 dB, which guarantees a high-quality imaging. 
The hybrid with the 5 mm aperture and $f_{0} = 5 mm$ also demonstrates a very good performance with slightly lower PSNR values. For the two other cases: $D=7, f_{0} = 7$ mm and $D=9, f_{0} = 7$ mm, we can see a much worse performance with PSNR values lower from 5 to 10 dB as compared with the best ones.


\begin{figure}[t!]
	\centering
	\includegraphics[width=0.6\linewidth]{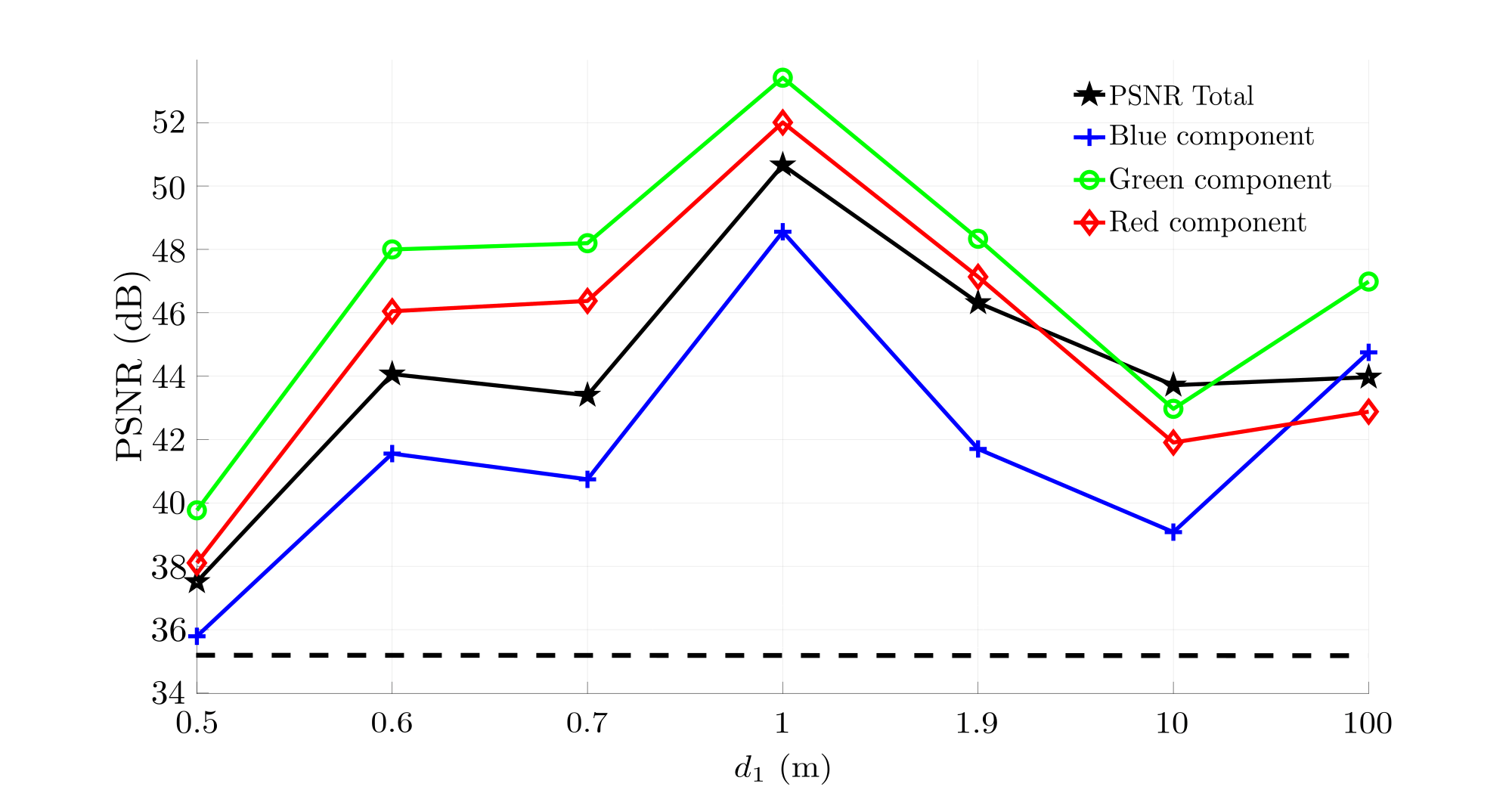}
	\caption{ The spectral performance of the best-optimized hybrid system ($D=6 mm$ and $f_{0}=5 mm$) is characterized by PSNRs calculated for the RGB channels as functions of $d_1$.  All curves are above the good imaging quality line 35 dB. The curves for color components mainly follow the behavior of the total PSNR curve (black). }
	\label{fig:rgb_PSNR}
\end{figure}
\begin{figure}[t!]
	\centering
	\includegraphics[width=0.7\linewidth]{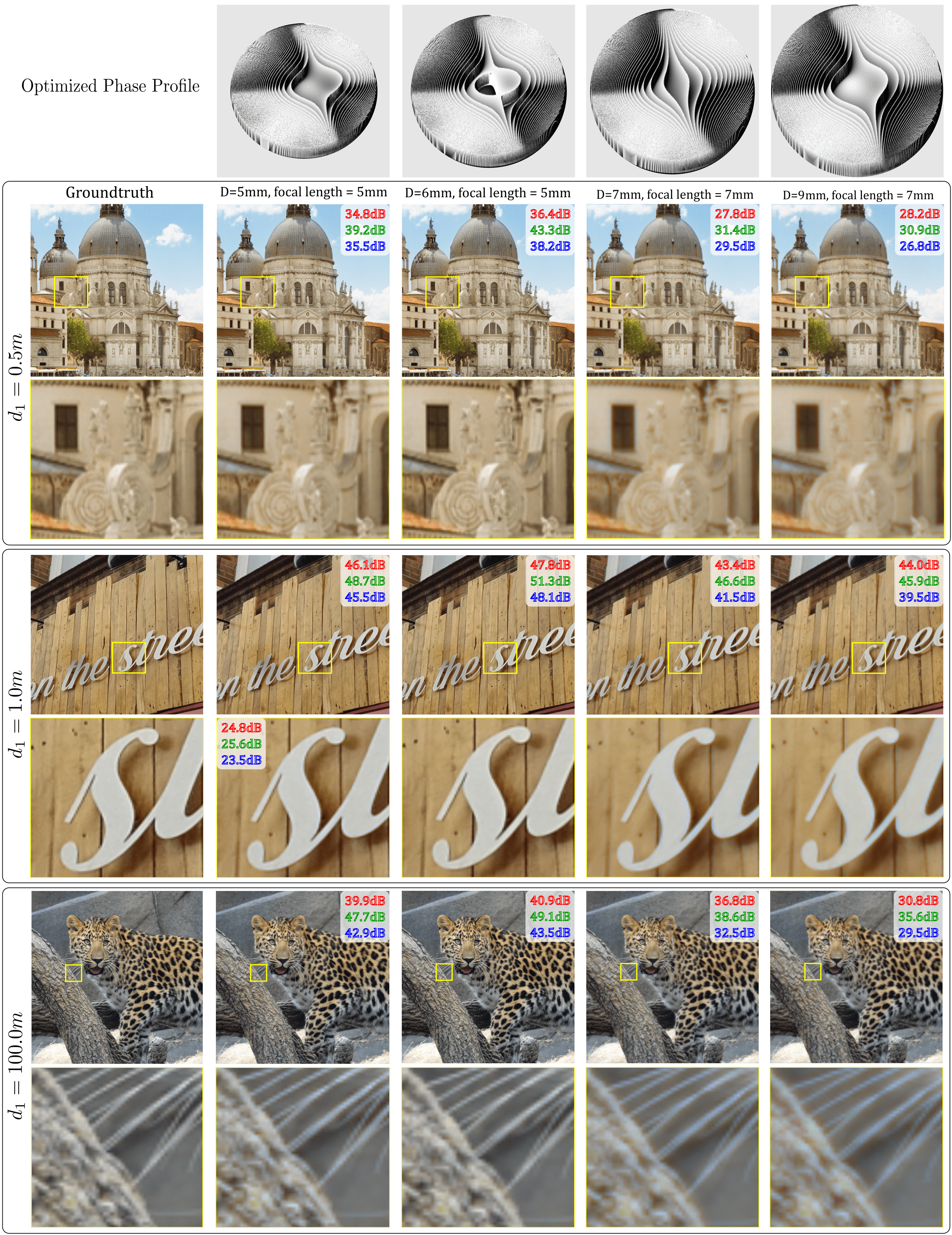}
	\caption{ Visual performance of the designed hybrid systems is illustrated for different diameters ($D=5,6,7,9$) mm with the optimal lens focal length as defined in Table \ref{tab:optics_compare}. The reconstructed images and their small fragments are shown for the distances $d_1 = (0.5, 1.0, 100.0) $ m. The color channels PSNR values are shown in these images. Thus, the comparison can be produced visually and numerically. The high-quality imaging for different colors and depths is achieved by the optical hybrid setups with the 6 and 5 mm diameter and 5mm focal length (columns 2 and 3). In contrast, the results for 7 and 9 mm diameters (columns 4 and 5) are suffering from strong chromatic aberration and the performance is degrading especially for off-focus distances $d_1 = 0.5$ and $d_1 = 100.0$ m. The optimized phase profiles of MPMs are shown in this first row of the image.}
	\label{fig:simulated}
\end{figure}

The spectral performance of the best-optimized hybrid system ($D=6 mm$ and $f_{0}=5 mm$) characterized by PSNRs calculated for the RGB channels as functions of $d_1$ is presented in Figure \ref{fig:rgb_PSNR}.  These curves with $PSNRs$  averaged over 200 test-images show the accuracy of imaging for each color channel and depth $d_1$. The $PSNR_{total}$, black curve in Figure \ref{fig:rgb_PSNR}, shows the accuracy as function of $d_1$ calculated for the all spectral channels simultaneously as averaged over 200 test-images. The color channel curves mainly follow the behavior of $PSNR_{total}$.  All these spectral curves are well above the 35 dB line confirming high-accuracy imaging for all $d_1$ and all spectral channels.

Figure \ref{fig:simulated} illustrates a visual performance of the designed hybrid systems of different diameters ($D=5,6,7,9$) mm with the optimal lens focal length as defined in Table \ref{tab:optics_compare}.
The reconstructed images and their small fragments are shown for the distances $d_1 = (0.5, 1.0, 100.0) $ m. The color channels PSNR values are shown in these images. Thus, the comparison can be produced visually and numerically. 
The optimized phase profiles of MPMs are shown in this first row of Figure \ref{fig:simulated}. 

Comparing these results, we may conclude, that the best results are achieved by the 5 mm and 6 mm diameter aperture sizes (columns 2 and 3) with an advantage of the latter one. For instance, for $d_1 = 0.5 m$, the improvement in PSNR is about 2 to 4 dB for color channels in favor of the hybrid optics with 6 mm lens diameter. Moreover, details and colors are better preserved in this case. This best setup provides uniformly better imaging quality for various depths and colors.
The zoomed fragments of the reconstructed images visually reveal clearly that the hybrid optics with 7 and 9 mm diameters (columns 4 and 5) are suffering from strong chromatic aberrations and quite blurry. 

\begin{figure}[t!]
	\centering
	\includegraphics[width=.6\linewidth]{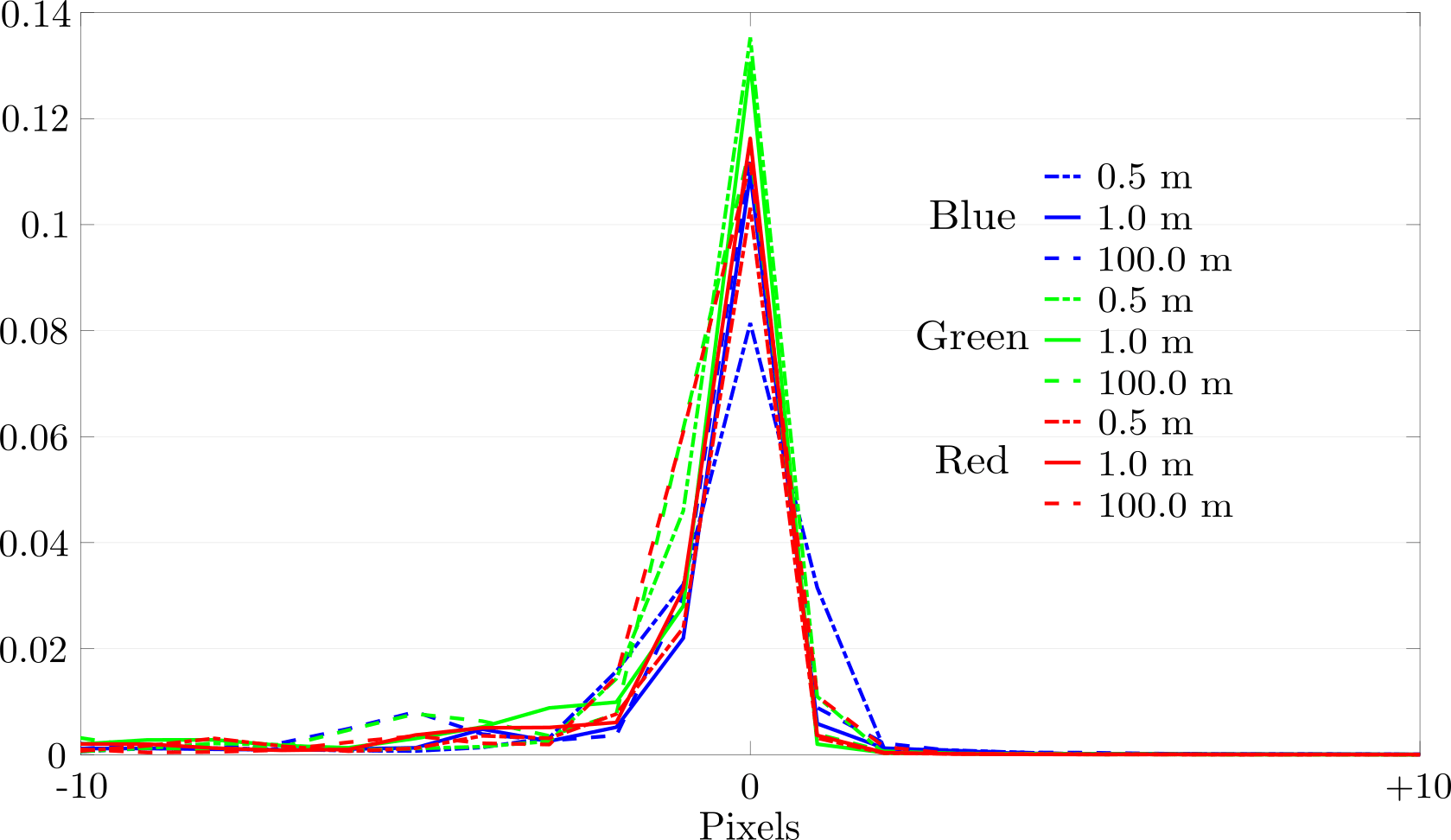}
	\caption{ For the best hybrid system ($D=6$ mm, $f_{0}=5$ mm), we show the cross-sections of the spectral PSFs for the three RGB channels and for the distances $d_{1}$  used in Figure \ref{fig:simulated}. These cross-section curves are well consolidated, what which explains a good performance of the imaging system for different distances $d_{1}$ and different color channels.}
	\label{fig:psfs}
\end{figure}

The advantage of the best hybrid optics with $D= 6 $ mm and $f_{0}=5$ mm is well seen as compared with its counterparts, what is in direct agreement with the results shown in Table \ref{tab:optics_compare}.
Additionally in Figure \ref{fig:psfs}, for this best hybrid system, we show the cross-sections of PSFs for the three RGB channels and for the distances $d_{1}$ used in Figure \ref{fig:simulated}. 
These cross-section curves are well consolidated, which explain a source of a good performance of the imaging system for different distances $d_{1}$ and different color channels. 

\section{Experimental Tests}
\label{experimental-result}

\subsection{Optical Setup and Equipment}

\begin{figure}[t!]
	\centering
	\includegraphics[width=0.7\linewidth]{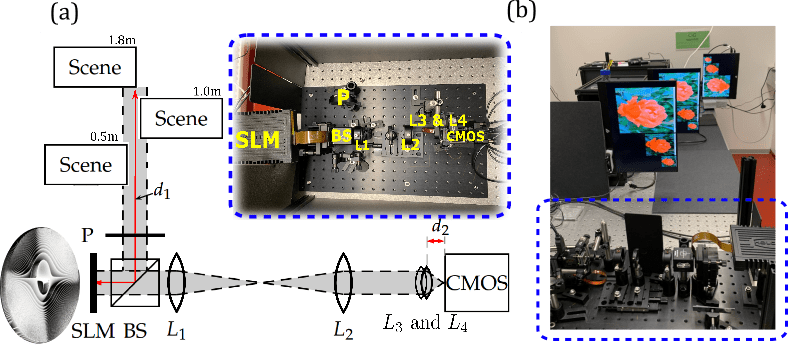}
	\caption{Experimental optical setup. Figure (a) illustrates the architecture of the optical setup with SLM and the photo of the corresponding hardware.  P is a polarizer, BS is a beamsplitter, SLM is a spatial light modulator. The lenses $L_1$ and $L_2$ form the $4f$-telescopic system projecting wavefront from the SLM plane to the imaging lenses $L_3$ and $L_4$, CMOS is a registering camera. $d_1$ is a distance between the scene and the plane of the hybrid optics ($L_3$ and $L_4$) and $d_2$ is a distance between this hybrid optics and the sensor. 
	Figure (b) shows the photo of this hardware with three monitors displaying the scenes (images) of three fixed distances $d_{1}=(.5,1.0.1.8)$ m. }
	\label{fig:setup}
\end{figure}

In this work, to implement our hybrid optics and in order to avoid building several MPM to physically analyze the performance of our camera, we build an optical setup based on a programmable
phase SLM to exploit its phase capabilities to investigate the performance of the designed hybrid setup.
 The optical setup is depicted in Figure~\ref{fig:setup}(a), where 'Scene' denotes objects under investigation; the polarizer, 'P', keeps the light polarization needed for a proper wavefront modulation by SLM; the beamsplitter, 'BS', governs SLM illumination and further light passing; the lenses '$L_1$' and '$L_2$' form a 4f-telescopic system transferring the light wavefront modified by SLM to the lenses '$L_3$' and '$L_4$' plane; the lenses '$L_3$' and '$L_4$' forms an image of the 'scene' on the imaging detector, 'CMOS'. 
 We use two lenses '$L_3$' and '$L_4$' tightly a fixed to each other  in order to get the hybrid's lens, as in Figure \ref{fig:system}, of a smaller focal length: $f_{0}=f_{1}/2$, where $f_{1}$ is the focal length of '$L_3$' and '$L_4$'.

For physical modeling of MPM phase delay, we use SLM: the Holoeye phase-only GAEA-2-vis SLM panel, resolution $4160\times2464$, pixel size $3.74~\mu$m.
'$L_1$' and '$L_2$' are achromatic doublet lenses with diameter $12.7$~mm and focal length $50$~mm;
Two BK7 glass lenses '$L_3$' and '$L_4$'are of diameter $6$~mm and focal length $10.0$~mm which results in $f_{0}=5.0$~mm;
'CMOS' Blackfly S board Level camera with the color pixel matrix Sony IMX264, $3.45~\mu$m pixel size and $2448\times2048$ pixels.
This SLM allows us to experimentally study the optical hybrid imaging with an arbitrary phase-delay distribution for the designed MPM. The MPM phase was created as an 8-bit \textit{*.bmp} file and imaged on SLM. We calibrated the SLM phase-delay response to the maximum value of $2.0\pi$ for wavelength equal to $510$~nm. This $2.0\pi$ corresponds to the value 255 of \textit{*.bmp} file for the phase-delay image of MPM. 

Figure \ref{fig:setup}(a) illustrates the architecture of the developed optical setup with SLM and the photo of the corresponding hardware.  Figure \ref{fig:setup}(b) shows the photo of this hardware with three monitors displaying the scenes (images) with three fixed distances $d_{1}=(0.5, 1.0, 1.8)$ m. The imaging monitors have a resolution of $1920\times1080$ and $570$ppi. The distance $d_{1}=1.0$ m is the focal point of the optical system.

\subsection{Optimization of image reconstruction provided a fixed MPM: learning-based approach}
This optimization is used in our physical experimental works provided that the optimized phase-delay profile of MPM with $D=6$ mm, obtained in the model-based approach,  is implemented by SLM. The outputs of the sensor are blurred images registered for a sequence of the train dataset images displayed on three monitors at three different depths, $d_1 = (0.5, 1.0, 1.8$ m). Convolutional Neural Network (CNN) is used to fit these blurred images to the known true target images. In this way, CNN designs the inverse imaging algorithm defined by the CNN parameters. For optimization, we exploit the stochastic gradient ADAM optimizer. The training process is running for 1244 high-quality images on three monitors which give in a total of 3732 registered images. The network has been trained for 320 epochs which takes two weeks on NVIDIA GeForce RTX 3090 GPU.

\begin{figure}[t]
	\centering
	\includegraphics[width=1.0\linewidth]{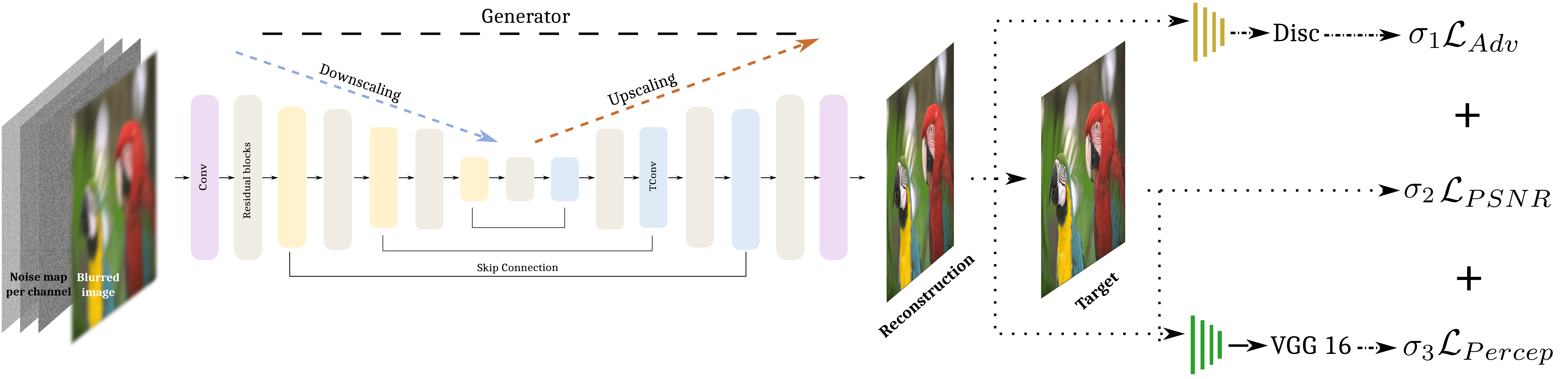}\vspace{.5em}
	\caption{ Inverse imaging UNet-based neural network architecture. The generator model is a U-net architecture that has seven scales with six consecutive downsampling and upsampling operations \cite{zhang2021plug}. We adopt a weighted combination of PSNR between estimated and ground truth images, $\mathcal{L}_{PSNR}$, and perceptual losses $\mathcal{L}_{Adv}$ and $\mathcal{L}_{Percep}$, with weights $\sigma_{1},\sigma_{2}$, and $\sigma_{3}$.}
	\label{fig:unet}
\end{figure}

Figure \ref{fig:unet} illustrates an architecture of CNN used in our experiments (DRUNet CNN  \cite{zhang2021plug}). We remark that this network has the ability to handle various noise levels for an RGB image, per channel, via a single model. The backbone of DRUNet is U-Net which consists of four scales. Each scale has an identity skip connection between $2\times 2$ strided convolution (SConv) downscaling and $2\times 2$ transposed convolution (TConv) upscaling operations. The number of channels in each layer from the first scale to the fourth scale are 64, 128, 256, and 512, respectively. Four successive residual blocks are adopted in the downscaling and upscaling of each scale. Each residual block only contains one ReLU activation function. The proposed DRUNet is bias-free, which means no bias is used in all the Conv, SConv and TConv layers \cite{zhang2021plug}.

An appropriate loss function is required to optimize the inverse imaging to provide the desired output. Thus, we use a weighted combination of PSNR between estimated and ground truth images ( $\mathcal{L}_{PSNR}$), perceptual loss, and adversarial loss which are given below.

\textbf{Perceptual loss: }To measure the semantic difference between the estimated output and the ground truth, we use a pre-trained VGG-16 \cite{simonyan2014very} model for our perceptual loss \cite{khan2020flatnet}. We extract feature maps between the second convolution (after activation) and second max pool layers $\varphi_{22}$, and between the third convolution (after activation) and the fourth max pool layers $\varphi_{43}$. Then, the loss $\mathcal{L}_{Percep}$ is the averaged PSNR between the outputs of these two activation functions for both estimated and ground truth images.

\textbf{Adversarial loss: } Adversarial loss \cite{goodfellow2014generative} was added to further bring the distribution of the reconstructed output close to those of the real images. Given the swish activation function \cite{ramachandran2017searching} as our discriminator $D$, this loss is given as $\mathcal{L}_{Adv} = -\log(D(I_{est}))$ where $I_{est}$ models the estimated image.

Our total loss for the proposed CNN inverse imaging while training is a weighted combination of these three losses and is given as, $\mathcal{L}_{CNN}=\sigma_{1}\mathcal{L}_{PSNR} + \sigma_{2}\mathcal{L}_{Percep} + \sigma_{3}\mathcal{L}_{Adv}$, where, $\sigma_{1},\sigma_{2}$ and $\sigma_{3}$ are empirical weights assigned to each loss. In this work, these constant are fixed as $\sigma_{1}=1.0, \sigma_{2}=0.6$, and $\sigma_{3}=0.1$. Lastly, the parameters of this networks to be optimized.

\begin{figure}[t!]
	\centering
	\includegraphics[width=0.7\linewidth]{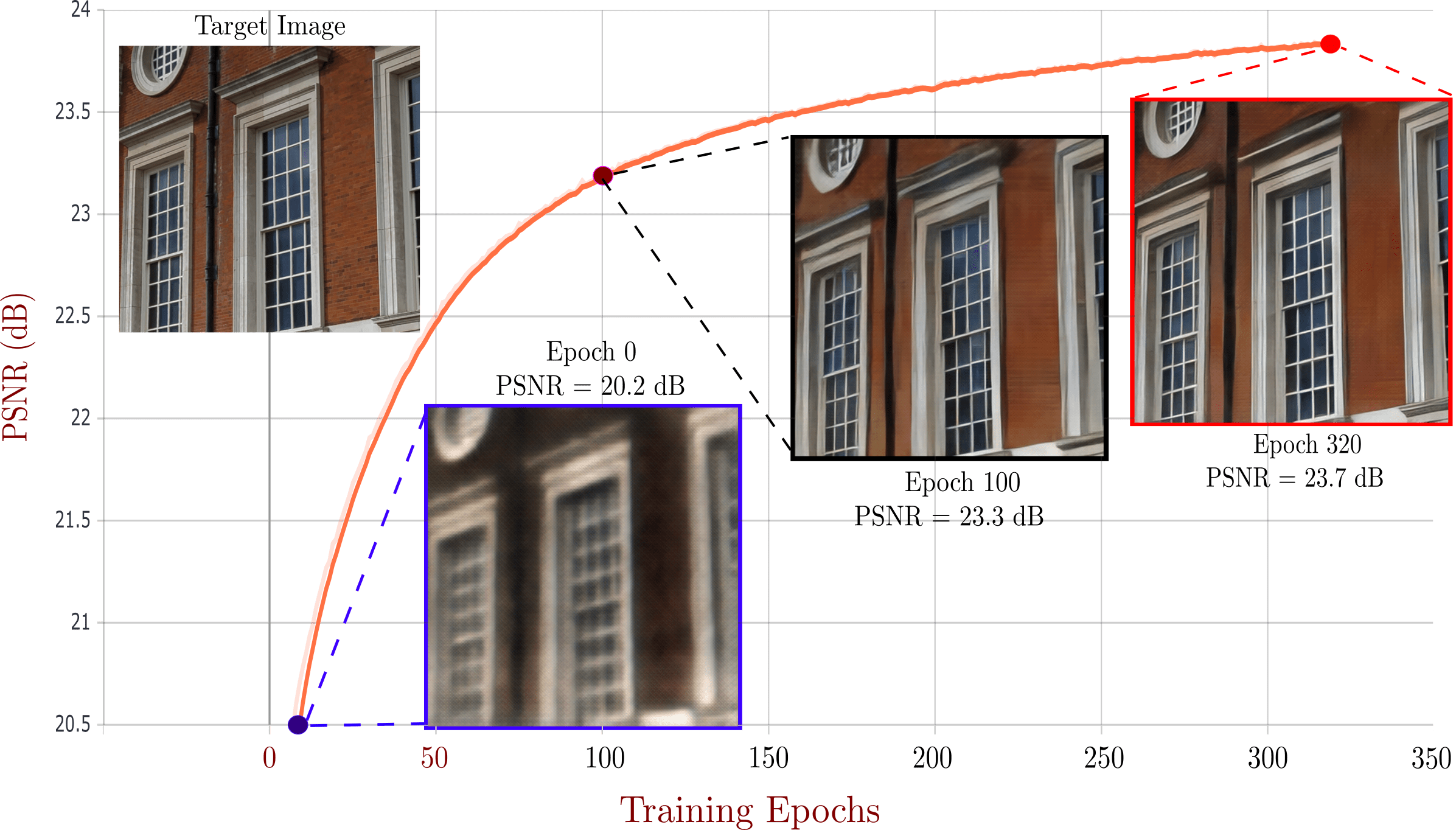}\vspace{.5em}
	\caption{ Performance of CNN for design of inverse imaging algorithm. The quality achieved by CNN starts from 20.5 dB and reaches 23.8 dB of PSNR for the training image set. The reconstructed images over the testing dataset are presented for three epochs 0, 100, and 320 for visualization of the training process. It could be seen that the trained network performs well for this task and the output image for epoch 320 is sharp enough. }
	\label{fig:training_quality}
\end{figure}

In Figure \ref{fig:training_quality} we report an evaluation of PSNR versus a number of epochs. From these results, we can see that the quality achieved by CNN for the designed hybrid system is quite high for the training data set. Illustrating reconstructed images chosen among the testing dataset are presented for epochs 0, 100, and 320. It could be seen that the trained network performs well and the output image for epoch 320 is sharp enough. The practical value of this approach to image processing design follows from using physical modeling of image formation including in particular wavefront propagation and mosaicing/demosaicing operations. 

\begin{figure}[t!]
	\centering
	\includegraphics[width=1.0\linewidth]{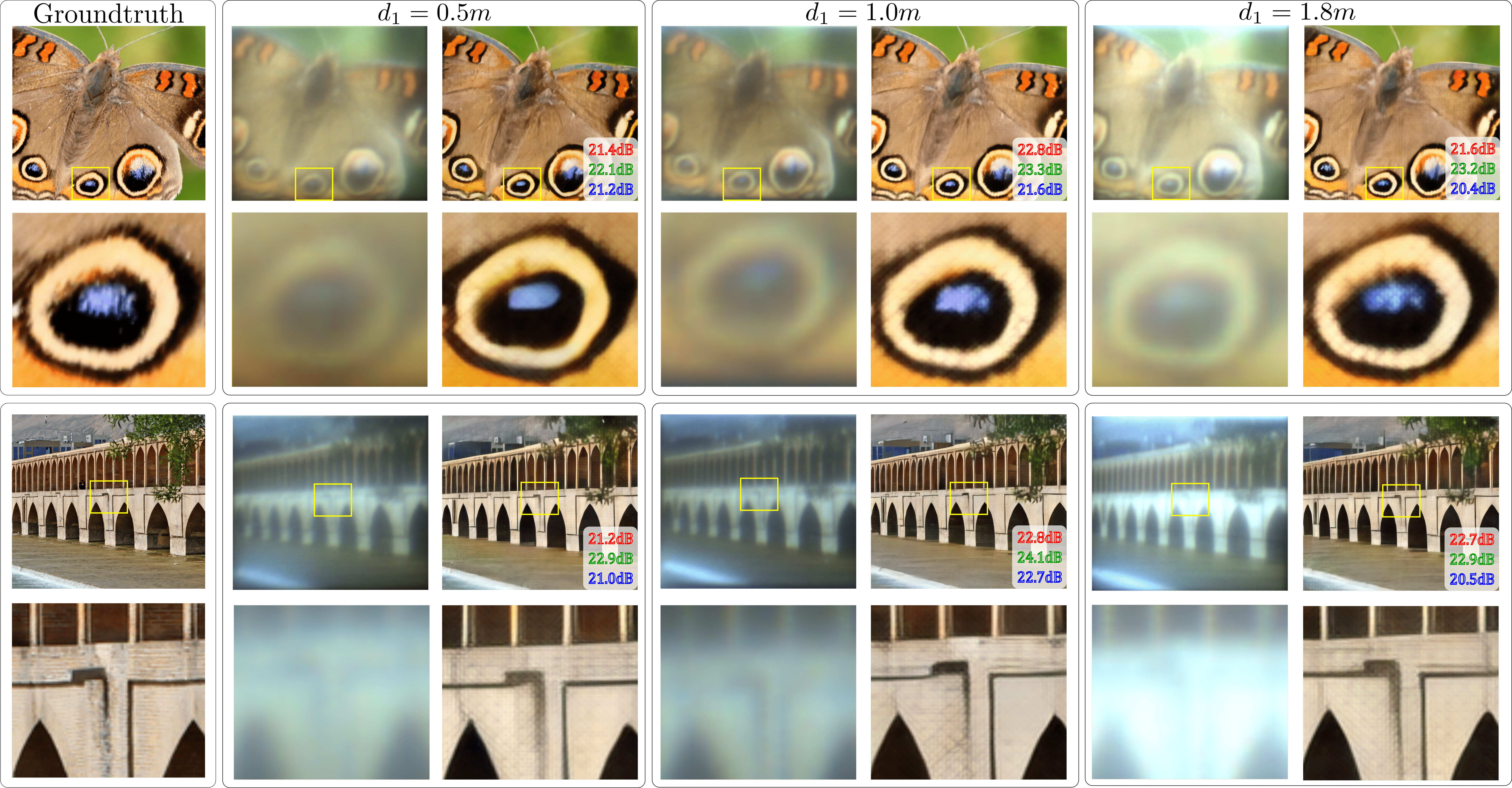}
	\caption{ Results for three monitor setup over the testing dataset. The reconstructed images with zoomed region at three different distances (imaging monitor-SLM): $d_1 = 0.5, 1.0, 1.8~ m$, for the optimized hybrid system. The PSNR values are reported for each depth and each color channel separately. The high-quality imaging with PSNR values of about 23 dB for different imaging depths and colors is achieved by the designed hybrid. }
	\label{fig:3monitor_result}
\end{figure}

\subsection{Experimental Results}
In this section, we present the results of two types of experiments. In the first one, the test-images are displayed on the three monitors as in Figure \ref{fig:setup}(b), the observations are blurred and the images are reconstructed by the trained CNN. These results are shown in Figure \ref{fig:3monitor_result}. 
In this scenario, we presented and evaluated the quality of reconstructions visually as well as numerically by PSNR values for each of the RGB color channels.

\begin{figure}[t!]
	\centering
	\includegraphics[width=0.65\linewidth]{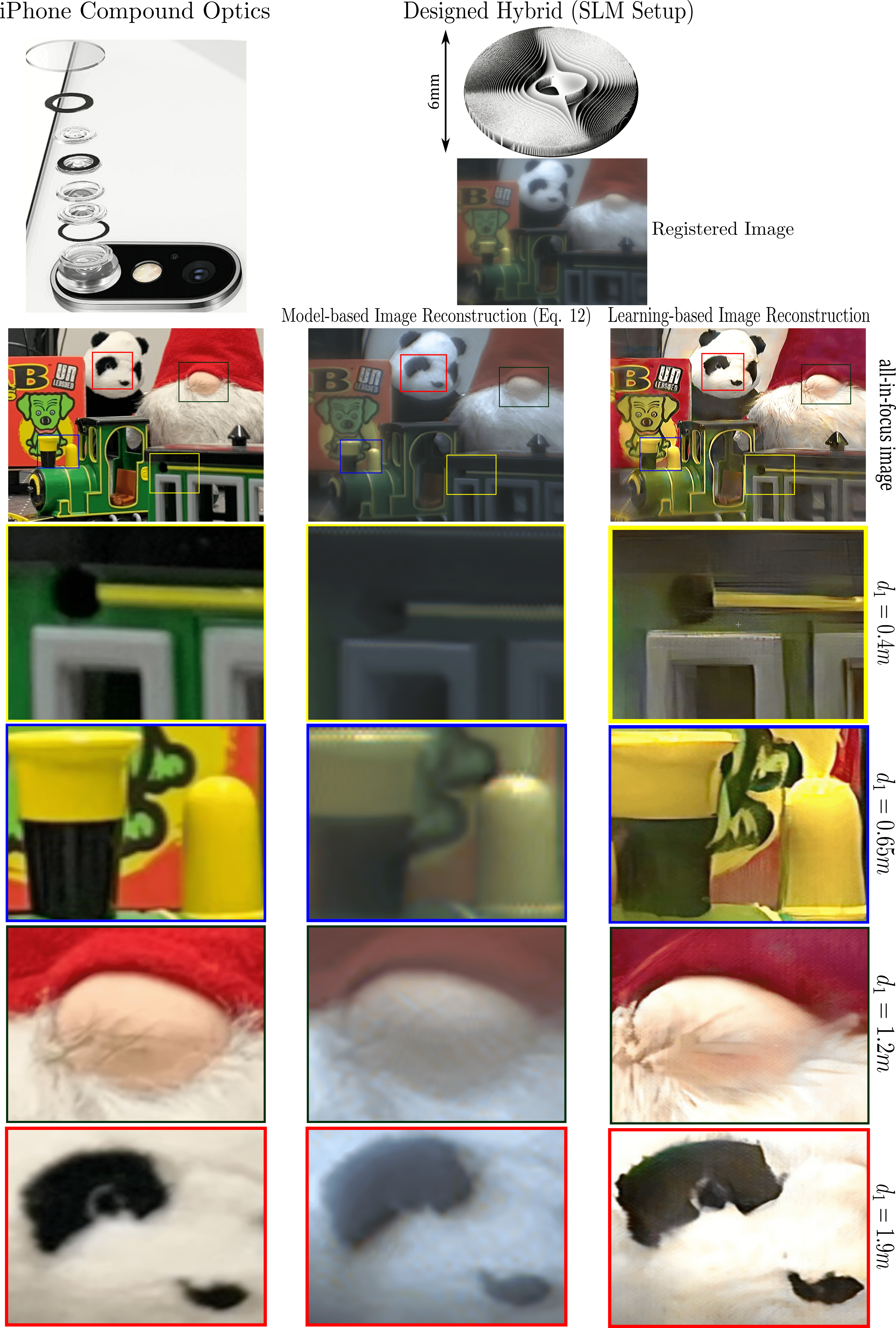}
	\caption{ Comparison of the designed hybrid diffractive imaging versus the compound lens camera of iPhone Xs Max. For the designed hybrid, two image reconstruction approaches are employed (columns 2 and 3) to recover the blurred image on the sensor: Model-based and learning-based. The obtained images are presented in row 3 with their enlarged fragments in rows 4, 5, 6, and 7 corresponding to four off-focus distances $d_1 = 0.4, 0.65, 1.2, 1.9~ m$, respectively. By comparing the results over the recovering approaches in the designed hybrid (columns 2 and 3), the advantage of using a deep UNet-style CNN is clear. For the iPhone camera, the imaging quality is not good for the close and far distances. The visual advantage in sharpness and color preservation is clearly in favor of the designed hybrid imaging. }
	\label{fig:experimental_result}
\end{figure}

In the second type of experiments, we image a scene composed  of different objects arbitrarily located  within the range (0.4-1.9)~m from the hybrid optics. This optical setup is used to evaluate the performance of the designed system in a real-world scenario for the EDoF imaging task. The performance of the designed
system is compared with the compound multi-lens commercial iPhone Xs Max camera. These results can be seen in Figure \ref{fig:experimental_result}. 

The results in Figure \ref{fig:3monitor_result} are presented in 7 columns for three depth distances: $d_1 = 0.5~ m$ (columns 2 and 3), $d_1 = 1.0~ m$ (columns 4 and 5), and $d_1 = 1.8~ m$ (columns 6 and 7). The Groundtruth column shows the true images.  
Two images from the test dataset are presented in this figure for comparison (rows 1 and 3) with one zoomed region (rows 2 and 4). We can see the zoomed fragments of the blurred noisy images on the sensor used for CNN image reconstruction as well as the corresponding reconstructed images. 
The zoomed sections for blurry and reconstructed
images visually reveal that the images are sharp and clear enough and the quality of imaging is high and more or less the same for different depths. Besides, the colors are well preserved properly along with distances.
If we compare the results numerically by PSNR, we could conclude that the PSNR values for different colors and depths are more or less the same at about 23 dB. It confirms that the designed hybrid imaging indeed demonstrates achromatic EDoF imaging.

The imaging results for the scene scenario are presented in Figure \ref{fig:experimental_result}. The scene consists of 5 objects located at different distances from $0.4$ ~m to $1.9 ~m$, approximately: $d_1 = 0.4 ~m$ (Train Wagon), $0.65~ m$ (locomotive), $1.15 ~m$ (ThorLab snack box), $1.2~ m$ (Dwarf Christmas Santa Claus Doll), and $1.9~ m$ (Panda toy). It is worth mentioning, that for iPhone (compound optics), we adjusted the focusing distance to $d_1 = 1.0$~ m as it is for the hybrid system.  The hybrid diffractive imaging is compared with imaging by the mobile phone camera iPhone Xs Max (column 1).

For the designed hybrid, two image reconstruction techniques are demonstrated:  model-based (column 2) and learning-based (column 3). In the model-based algorithm, the scene is reconstructed using the calculated color channel PSFs and the inverse imaging according to Eq. (\ref {misfocus_color_data101}). After this step, a denoising process equipped with a sharpening procedure \cite{Dabov07jointimage} is performed over the estimated scene to improve the quality of imaging. This final denoised image is returned as the estimated scene from experimental data.  Contrary to it, the learning-based inverse imaging uses the trained UNet.  For a detailed comparison, the four zoomed fragments of the images are shown in rows 4, 5, 6, and 7 which correspond to the scene's objects of different out-of-focus distances. Comparing columns 2 and 3, we may note an obvious advantage of the learning-based inverse imaging.  The model-based approach is not able to recover all details, the output image is still blurry, and chromatic aberrations are strong. There are a number of reasons for this advantage. First of all, it concerns a mismatch between reality and the analytical modeling of image formation by PSFs. Second, the mosaicing/demosaicing are not included in our modeling. The leaning-based approach allows successfully compensate  these drawbacks of the analytical modeling. 

Comparison of the learning-based hybrid imaging (column 3) versus the iPhone camera imaging (column 1) results in an exciting conclusion about a quite clear advantage of the hybrid diffractive imaging. This advantage is obvious for sharpness of images for all distances. Thus, hybrid imaging demonstrates high-quality all-in-focus imaging. Concerning color aberrations: red and green perhaps not be properly presented by the hybrid but the white color is definitely perfect. Thus overall, the hybrid diffractive imaging can be tread at least as quite competitive and even advanced with respect to the commercial iPhone with multi-lens optics. Here we need to note that the white balance and $\gamma$ correction procedure have been produced for the images reconstructed by the hybrid system in order to have a fair comparison with the iPhone camera.


\section{Conclusion}
\label{sec:conclusion}
For the first time, aperture size, lens focal length, and distance between MPM and sensor are considered as optimization variables for diffractive achromatic EDoF imaging. It is shown in this paper that the design and end-to-end optimization proposed for computational imaging with optics composed from a single refractive lens and a diffractive phase-encoded MPM is quite successful. In particular, comparison versus imaging by the multi-lens iPhone camera is definitely in favor of the designed imaging system. 
One of the novelties proposed and exploited in this paper is a the physical modeling of MPM by SLM which allows an on-line design of free-shape phase encoding for diffractive optics excluding a mismatch of theoretical image formation modeling and physical reality.  As further work, we consider three lines of development. First, an improved design of SLM phase delay based on the Hardware-In-the-Loop approach.
Second, a design of 'thick' MPMs (Fresnel order much larger than 1) is not possible using SLM but allows to get more efficient phase encoding and imaging. Third, an implementation of MPM as a physical phase mask.

\acknowledgments 
 This work is supported by the CIWIL project funded by Jane and Aatos Erkko Foundation, Finland. \vspace{-0.6em}

\bibliography{report} 
\bibliographystyle{spiebib} 

\end{document}